\documentclass[12pt]{article}
\usepackage{amsmath,bm}
\usepackage{amssymb}
\usepackage{graphicx}
\usepackage{amsfonts}
\usepackage{slashed}

\renewcommand{\baselinestretch}{1.2}
\let\non\nonumber

\global\newcount\itemno \global\itemno=0

\def\itemaut#1{\global\advance\itemno by1\noindent\item{\the\itemno.}#1}

\newif{\ifeq}           
\eqtrue                 

\newcommand{\be}{\begin{equation}}
\newcommand{\ee}{\end{equation}}
\newcommand{\bes}{\begin{equation*}}
\newcommand{\ees}{\end{equation*}}
\newcommand{\bea}{\begin{eqnarray}}
\newcommand{\eea}{\end{eqnarray}}
\newcommand{\bean}{\begin{eqnarray*}}
\newcommand{\eean}{\end{eqnarray*}}

\newcommand{\scz}{\setcounter{equation}{0}}

\def\({\left(}
\def\){\right)}
\def\[{\left[}
\def\]{\right]}

\def\frac#1#2{{#1 \over #2}}

\renewcommand{\a}{\alpha}
\renewcommand{\b}{\beta}
\renewcommand{\d}{\delta}
\newcommand{\g}{\gamma}

\renewcommand{\r}{\rho}
\newcommand{\s}{\sigma}

\renewcommand{\l}{\lambda}

\newcommand{\w}{\omega}
\newcommand{\m}{\mu}
\newcommand{\n}{\nu}

\newcommand{\eps}{\epsilon}

\renewcommand{\t}{\theta}

\newcommand{\phib}{\bar{\phi}}

\newcommand{\z}{\xi}

\newcommand{\CL}{{\cal L}}

\newcommand{\CN}{{\cal N}}
\newcommand{\CO}{{\cal O}}

\newcommand{\IZ}{{\mathbb Z}}

\def\xv{\vec{x}}

\def\eg{{\it e.g.}}
\def\ie{{\it i.e.}}

\def\etal{{\it et. al.}}

\newcommand{\lsim}{\,\raise.3ex\hbox{$<$\kern-.75em\lower1ex\hbox{$\sim$}}\,}
\newcommand{\gsim}{\,\raise.3ex\hbox{$>$\kern-.75em\lower1ex\hbox{$\sim$}}\,}

\def\p{\partial}

\def\II{\relax{I\kern-.10em I}}

\def\sch#1#2{$Sch^{#1}_{#2}$}
\def\lif#1#2{$Lif^{#1}_{#2}$}
\def\bDLCQ{$\b$DLCQ}

\begin{document}
\begin{titlepage}
\begin{flushright}
MIT-CTP/4001
\end{flushright}

\vspace{1cm}
\begin{center}
{\Large \bf 1/$N$  Effects in Non-Relativistic Gauge-Gravity Duality}\\
\end{center}

\vspace{.4cm}
\begin{center}
{Allan Adams$^a$, Alexander Maloney$^{b}$, Aninda Sinha$^c$ and Samuel E. V\'azquez$^c$}\\
\end{center}

\vspace{.3cm}
\begin{center}
{\em $^a$ Center for Theoretical Physics, MIT, Cambridge, Massachusetts, 02139, USA}\\
{\em $^b$ McGill Physics Department, 3600 rue University, Montreal, QC H3A 2T8, Canada}\\
{\em $^c$ Perimeter Institute for Theoretical Physics, Waterloo, Ontario N2L 2Y5, Canada}
\end{center}

\vspace{8mm}
\begin{center}
{\bf Abstract}
\end{center}
\noindent
We argue that higher-curvature terms in the gravitational Lagrangian lead, 
via non-relativistic gauge-gravity duality, to finite renormalization of the 
dynamical exponent of the dual conformal field theory.  Our argument 
includes a proof of the non-renormalization of the Schr\"odinger and 
Lifshitz metrics beyond rescalings of their parameters, directly generalizing 
the AdS case.  We use this effect to construct string-theory duals of 
non-relativistic critical systems with non-integer dynamical exponents, then
use these duals to predict the viscosity/entropy ratios of these systems. 
The predicted values weakly violate the KSS bound.

\vfill
\begin{flushleft}
\today
\end{flushleft}

\vfil
\end{titlepage}
\newpage

\renewcommand{\baselinestretch}{1.1}  
\renewcommand{\arraystretch}{1.5}

\section{Introduction}\scz

The study of non-relativistic conformal field theories has accelerated considerably in recent years, both theoretically \cite{NRCFTth}\ and experimentally \cite{NRCFTex}.  Since these systems are typically strongly coupled, theoretical progress has depended largely on sophisticated non-perturbative techniques or large-$N$ toy models (for example, the $\eps$ \cite{eps} or large-$N$ \cite{N} expansions).  A general and computationally effective strong-coupling description would be extremely useful. 

It has recently been argued that many non-relativistic CFTs  (NRCFTs) should admit dual strong-coupling descriptions as gravitational systems.  In this duality, the non-relativistic conformal group is realized as the isometry group of a dual spacetime geometry \cite{original}.  A special subset of these geometries were subsequently embedded in string theory (which allowed a precise identification of the dual NRCFT) and warmed to finite temperature by the heat of the Hawking fire \cite{malda,Adams,Herzog}.

Thus far, only a special subset of NRCFTs  (those with Schr\"odinger symmetry with dynamical exponent $z=2,4$) have found such a home in string theory\footnote{Recently, Hartnoll and Yoshida \cite{HartnollYoshida}\ introduced a novel and beautiful embedding of certain $z\neq2$ spacetimes into string theory.  However, due to the more intricate form of these metrics (explicitly, the Schr\"odinger part of the metric varies non-trivially over the compact Sasaki-Einstein space), it is not clear how to use these spacetimes to construct a dual CFT.  It would be very interesting to illuminate this point.}.  Meanwhile, many notable effects discovered in the study of NRCFTs -- for example, the renormalization of the dynamical exponent -- remain obscure.  Finally, with only a few known examples, it is unclear what features of non-relativistic gauge/gravity duality are robust, or even universal.  For example, do all NRCFTs satisfy the KSS bound, $\eta/s \geq 1/4\pi$?  If not, does some quantum version persist, as in the relativistic case, or are these systems even more non-universal?  What about more general hydrodynamic physics, much of which is not universal even in the relativistic case?

In this paper we argue that $R^{2}$ corrections to the gravitational action map, via gauge-gravity duality, to a renormalization of the dual NRCFT away from the naive classical result.  By engineering string configurations in which suitable higher-curvature terms are under control, we construct explicit stringy NRCFT-gravity dualities with non-integral dynamical exponents $z\neq2$.  For example, we find that the NRCFT defined by the \bDLCQ\ of an $\CN=2$ gauge theory with gauge group $Sp(N)$ is dual to type IIB string theory on \sch{z_{N}}{5}. The dynamical exponent $z_{N}$ of the strongly-coupled theory has a large-$N$ expansion\footnote{This expansion is reminiscent of the large-$N$ renormalization of exponents found in other NRCFTs \cite{N}.  Reproducing our expansion directly within the dual field theory using such techniques would provide a strong test of non-relativistic gauge-gravity duality.},
\be
z_{N}=2+{2\over 27 N} +\CO\({1\over N^{2}}\)\;.
\ee
Applying various lessons from gauge-gravity duality, we use these results to argue that the KSS bound may well be violated by quantum effects in NRCFTs as it is in the relativistic case \cite{kp,Steve}, with the viscosity to entropy ratio for our main example given by,
\be
{\eta\over s} = {1\over 4\pi}\(1-{1\over 2N}\) .
\ee
This weak quantum violation of the KSS bound suggests that typical NRCFTs, whose critical exponents are generically renormalized, will not universally satisfy the KSS bound.

This paper is organized as follows.  In Section 2 we review the geometry of Schr\"odinger space and examine the conditions under which a general theory of gravity+matter will enjoy Schr\"odinger solutions to its equations of motion.  We then engineer explicit examples of higher-curvature theories of gravity supporting Schr\"odinger space.  We also repeat these arguments and constructions for metrics with Lifshitz, rather than Schr\"odinger, isometry group.  In Section 3 we study the effects of higher-curvature terms present in the IIB action.  Treating these deformations perturbatively, we find new solutions of the IIB with non-integer dynamical exponent, $z\neq2$; for special values of the higher-curvature corrections, we argue that the dual CFT is the non-relativistic \bDLCQ\ of a $Sp(N)$ gauge theory.  Again, similar results are found for Lifshitz spacetimes.  In Section 4 we comment on the hydrodynamics of these general-$z$ systems and argue that they weakly violate the KSS bound.  We close in Section 5 with comments and open questions.

\section{Non-Renormalization of Schr\"odinger Space}\scz

It is a familiar and beautiful fact that quantum corrections to the gravity+matter action do not lift AdS solutions of the equations of motion -- all they can do is renormalize the AdS radius in units of $G_{N}$.  In this section we will prove a similar result for non-relativistic Schr\"odinger and Lifshitz metrics, with the minor modification that both the radius of curvature and the dynamical exponent $z$ may be renormalized.  We begin by describing the geometry of Schr\"odinger metrics.

\subsection{The Non-Relativistic Conformal Algebra}
The non-relativistic conformal algebra in $d$ spatial dimensions, known as the $d$-dimensional Schr\"odinger algebra, is generated by spatial translations, $P_{i}$, rotations $M_{ij}$, boosts $K_{i}$, and time translations, $H$,  together with a dilatation operator, $D$, and a ``number'' operator, $N$, with non-trivial commutation relations,
\bea
[D,P_{i}] = iP_{i}  ~~~~~~~~~~ [D,K_{i}]=&&\hspace{-6mm} i(1-z)K_{i}  ~~~~~~~~~~ [D,H]=izH \non \\
\[D,N\] = i ( 2-z) N  ~~~~&&\hspace{-6mm}~~~~ [P_{i},K_{j}]=-i\delta_{ij}N
\eea
The dynamical exponent $z$ determines the relative scaling between the time and spatial coordinates, $[t]$=$[x]^{z}$, or equivalently fixes the dispersion relation to, $\w \sim k^{z}$.  The algebra imposes no constraints on the value of $z$ other than reality.  We thus have a one-parameter family of conformal algebras labeled by $z$.

It is useful to emphasize how different this is from the relativistic case.  Part of the power of the relativistic conformal group is its rigidity -- there are no free parameters that could possibly depend on the dynamics.  Relativistic conformal invariance can thus be used to powerfully constrain dynamical effects.  In the Schr\"odinger case, however, we have a 1-parameter family of conformal algebras, with the dynamical exponent $z$ providing some information about how the dynamics of the particular system realizes the conformal symmetry.  The weak-coupling symmetries of the system are simply insufficient to completely fix the conformal structure in the NR case.  Indeed, it is known that $z$ may be shifted away from its naive classical value by interactions \cite{N}, an effect we shall examine in detail in what follows.

This is not to say all values of $z$ are equivalent: physically, different values of $z$ correspond to different dispersion relations;  algebraically, a shift of $z$ cannot be realized by any operator inside the algebra.  Shifting $z$ is not a symmetry of an NRCFT.  Meanwhile, and importantly, for special values of $z$ the algebra may be extended to a larger algebra (though of course the full extended symmetry need not be realized).    

For example, consider the special case $z=2$. The dispersion relation is $\w\sim k^{2}$, so this is a conformal version of the familiar Galilean group.  This case has two important special properties.  First, the generator $N$ is central, since $[D, N]=0$.  Representations of the $z=2$ algebra are thus labeled by two numbers, a dimension $\Delta$ and a ``number'' $\ell$.  (For fermions at unitarity, $N=\psi^{\dagger}\psi$, so $\ell$ is precisely the fermion number.)  Second, the $z=2$ algebra may be extended by a ``special conformal'' generator $C$ with non-trivial commutators
\be
[D,C]=-2iC\;,\;\;\; [H,C]=-iD\;.
\ee

A more familiar special case is $z=1$.  This corresponds to the relativistic dispersion relation $\w\sim k$, with $H$, $N$ and $P_i$ scaling uniformly under dilatations.  The algebra may thus be extended by operators generating rotations and boosts mixing the number operator with the time and spatial directions.  This enhanced algebra is the just the usual relativistic conformal algebra, $SO(d+1,2)$, aka the AdS isometry group.

Since these extended algebras are larger than the algebra with other values of $z$, systems realizing these extended Schr\"odinger conformal algebras are considerably more tightly constrained than those with general $z$.  Explicitly, since the conformal structure has no free parameters to depend on the details of the dynamics, these systems are considerably more robust to quantum modification than theories with more general $z$: were a radiative effect to shift $z$, the extended symmetry would be broken!  By contrast, for systems respecting the non-extended conformal symmetry, $z$ may be consistently renormalized, with the dispersion relation for low-lying modes determined by the details of the conformal fixed point.

\subsection{Schr\"odinger Spacetimes}
As in the case of the relativistic conformal group, the non-relativistic conformal group may be realized as the isometry group of an associated spacetime.  We'll denote this spacetime by \sch{z}{d+3}, where $d$ counts the number of spatial dimensions and $z$ is the dynamical exponent.  The metric on \sch{z}{d+3} is
\be \label{SchMetric}
ds^2 = L^2 \left( - {dt^2 \over r^{2z}} + {-2 dt d\xi + d\xv^2 + dr^2\over r^{2}}\right) ,
\ee
where $x^i$, $i=1,\dots\ d$ label the spatial coordinates and $L$ controls the overall radius of curvature; when we wish to denote the dependence on $L$ explicitly, we will write \sch{z, L}{d+3}.  This metric has isometries generated by the following Killing vectors
\bea\label{killings}
\non
M_{ij} = -i( x^i \p_j - x^j \p_i ),~~~~~
P_i&&\hspace{-6mm}=-i \p_i,~~~~~
H = -i \p_t,~~~~~
K_i = -i (-t \p_i+ x^i\p_\z)\;, \\
D = -i (zt\p_t + x^i \p_i &&\hspace{-6mm}+ (2-z)\z \p_\z + r\p_r),~~~~~
N = -i \p_\z . 
\eea
It is straightforward to verify that these obey the Schr\"odinger algebra.  It is not difficult to exponentiate these infinitesimal generators to find the action of the Schr\"dinger group on the spacetime (\ref{SchMetric}).
The special case \sch{z=2}{d+3}\ enjoys the additional Killing vector,
\be
C = -i (t^2 \p_t + t x^i \p_i - {\xv^2 + r^2\over 2}\p_\z + r t\p_r),
\ee
realizing the special conformal extension of the $z=2$ algebra. The goal of the rest of this section is to study the conditions under which \sch{z}{d+3} may be embedded in a generic theory of gravity; we apply these considerations to IIB string theory in the next section.

\subsection{Invariant two-forms in Schr\"odinger Spacetimes}
Suppose we are looking for \sch{z}{d+3} solutions to the equations of motion of gravity coupled to some matter sector.  The basic objects in these equations -- the stress tensor $T_{\mu\nu}$ and the Einstein tensor $G_{\m\n}$ evaluated on (\ref{SchMetric}) -- are symmetric two tensors invariant under the Schr\"odinger symmetries (\ref{killings}).  Due to the symmetry of this geometry, however, there are very few such symmetric invariant two-forms.  Expanding the Einstein equation in a basis of such symmetric invariant two-forms will allow us to reduce the search for solutions to a simple algebraic form.  With this motivation, we now proceed to classify all symmetric two-forms invariant under the Schr\"odinger symmetries.
 
Let $\tau = \tau_{\mu\nu}dx^\mu dx^\nu$ be a symmetric two-tensor invariant under the Schr\"odinger group, \ie
\be
\CL_v \tau_{\mu\nu} = 0\;,
\ee
for all of the Killing vectors $v$ listed above.  Invariance under $H$, $N$ and $P_i$ implies that the components of $\tau_{\mu\nu}$ are functions of $r$ only.  Invariance under $M_{ij}$ implies that $\tau_{ti}, \tau_{\z i}, \tau_{ri}$ vanish and that $\tau_{ij}$ is proportional to $\delta_{ij}$.   Invariance under $K_i$ implies that $ \tau_{ \z \z }$ and $\tau_{\z r}$ vanish and that $\tau_{t\z} = \tau_{ii}$.  Finally, invariance under $D$ implies that
\be
\tau ~=~
\a  \[{dt^2\over r^{2z}}\] ~+~
\b  \[{-2 dt d\z + d\xv^2\over r^2}\] ~+~
\g  \[{dr^2 \over r^2}\] ~+~
\d  \[{dt dr \over r^{z+1}}\]\;,
\ee
for some constants $\alpha,\beta,\gamma, \delta$.  Thus for generic values of $z$ there is a four parameter family of symmetric tensors invariant under the continuous symmetries of \sch{z}{d+3}. The stress tensor must in addition be conserved,
\be
\nabla^\mu \tau_{\mu\nu} =0,
\ee
which restricts the family of allowed stress tensors to the two parameter class
\be\label{allowed}
\tau = \a  \[{dt^2\over r^{2z}}\] ~+~
\b  \[{-2 dt d\z + d\xv^2+dr^2 \over r^2}\].
\ee
In the special case $z=2$, one may show that this two parameter family is in addition invariant under $C$.
We thus learn that any conserved, Schr\"odinger-invariant symmetric rank two tensor is completely specified by the two constants $\a$ and $\b$ defined above.

This observation allows us to discuss the construction of \sch{z}{d+3} solutions under general circumstances.  Consider the equations of motion of gravity coupled to an arbitrary matter sector via the action
\be
S = {1\over 16 \pi G} \int \sqrt{g} (R -2 \Lambda)+ S_m(g,\phi_i)\;.
\ee
Here $S_m$ denotes an arbitrary matter sector with matter fields $\phi_i$ where $i=1,\dots,N$.  We will also allow $S_m$ to include higher curvature corrections to the
Einstein-Hilbert action.
The metric equation of motion is
\be\label{meteq}
R_{\mu\nu} + \left(\Lambda - {1\over 2}R \right) g_{\mu\nu} = -8 \pi G T_{\mu\nu}\;,
\ee
where
$T_{\mu\nu} = {\delta S_m / \delta g^{\mu\nu}}$ includes the usual matter stress tensor as well as additional functions of the Riemann tensor due to the presence of the higher curvature terms.

Let us now evaluate this equation on the \sch{z}{d+3} metric (\ref{SchMetric}).  The left hand side is automatically a conserved two tensor which is invariant under the \sch{z}{d+3} isometries, which may thus be written in the form (\ref{allowed}). It is not hard to check that the $\a$ and $\b$ coefficients are simple functions of $z$ and $L$,
\bea \label{LHScoef}
\a^{G} &=& (z-1)(2z+d-3)-{(d-1)(d-2)\over 2} -\Lambda \;,\\
\b^{G} &=& {(d-1)(d-2)\over 2}  +\Lambda\;\nonumber.
\eea
Since the left hand side of our metric equation (\ref{meteq}) is a Schr\"odinger-invariant symmetric two-tensor, the right hand side must also take this form. Hence it must fall in the classification of symmetric tensors described above, and in particular must have the values of $\a$ and $\b$ given in (\ref{LHScoef}).

The invariance of $T_{\mu\nu}$ is a non-trivial constraint on the $\phi_i$.  The simplest way to impose this constraint is to demand that the fields themselves are \sch{z}{d+3} invariant
\be
\CL_v \phi_i = 0\;.
\ee
Indeed, this constraint is nothing more than the statement that the full background field configuration -- rather than just the metric -- is \sch{z}{d+3} invariant.  In this case, $T_{\mu\nu}$ is automatically \sch{z}{d+3} invariant.
If $\phi_i$ is a field (or a collection of fields) of spin $0$ or $1$ it is straightforward to check that \sch{z}{d+3} invariance implies that the $\phi_i$ are just constants.   We may then view the action as a function of the constants $\phi_i$.
The $\phi_i$ equations of motion are just minimization with respect to $\phi_i$
\be
{\partial S_m (g^{sch},\phi^i) \over \partial \phi_i} =0\;.
\ee
These may be regarded as equations for the unknown constants $\phi_i$.  Since these are $N$ equations for $N$ unknowns, the system will ``generically" have a solution.  More precisely, we have demonstrated that if we have an action which has a \sch{z,L}{d+3} solution, then for sufficiently small variations of the couplings in the matter sector \sch{z,L}{d+3} will remain a solution to the equations of motion.  The values of $\phi_i$ will change as we vary the parameters of the action.

Of course, it is not strictly necessary that the background fields $\phi_i$ are \sch{z}{d+3} invariant in order for $T_{\mu\nu}$ to be invariant.  We will however demand that any gauge invariant observables must be invariant under the full \sch{z}{d+3} symmetry.  For example, if we have a $U(1)$ gauge field $A_\mu$, then we should in general allow non-constant configurations, subject to the constraint that the fieldstrength $F = dA$ is \sch{z}{d+3} invariant.  In this case there is a one-parameter family of such allowed fieldstrengths
\be
F = e {dt \wedge dr \over r^{z+1}}\;,
\ee
Of course in this case the stress tensor depends only on the constant $e$ and not on the gauge dependent form of $A_\mu$.

For any such matter sector, then, the stress tensor evaluated on \sch{z,L}{d+3} takes the form,
\be
T_{\mu\nu} = \a(L,z,\phi_i) ~\! \[{dt^2\over r^{2z}}\] +
\b(L,z,\phi_i) ~\! \[{-2 dt d\z + d\xv^2+dr^2 \over r^2}\]\;,
\ee
where $\a,\b$ are functions of the curvature radius $L$, the dynamical exponent $z$ and the matter fields $\phi_i$.  The specific form of these functions will depend on the parameters of the matter action $S_m$.
The metric equations thus reduce to,
\be
\alpha^G = \alpha(L,z,\phi_i), ~~~~~~ \beta^G = \beta(L,z,\phi_i)\;.
\ee
As the values of $\phi_i$ -- or at least of the gauge invariant observables on which the stress tensor depends -- are fixed by the $\phi$ equations of motion, these may be viewed as two equations for the two unknowns $L$ and $z$.  

We conclude that \sch{z,L}{d+3} metrics arise generically as solutions of gravity coupled to sufficiently general matter -- one does not need to  tune the parameters in the action to find a \sch{z,L}{d+3} solution.   In particular, if \sch{z,L}{d+3} solves the equations of motion for one action, then for any small deformation of the parameters in the action we can find a new solution \sch{z',L'}{d+3} with nearby values of $z$ and $L$.    Conversely, variation of the action (due for example to quantum corrections) can only lead to renormalization of the parameters $z$ and $L$ appearing in the metric.  Thus as quantum corrections are included in the effective action, the Schr\"odinger symmetry of the spacetime will be bent (by changing the value of $z$) but not broken.

In many cases of interest we may wish to ask whether a particular value of $z$ is found as a solution to the equations of motion.  In this case we are required to impose a one parameter constraint on the parameters appearing in the action $S_m$.
The value of $L$ is then given by simply minimizing the action on \sch{z,L}{d+3} as a function of $L$ i.e. by solving
\be
{d S(L) \over  dL} = 0\;.
\ee
This is nothing more than the Schr\"odinger version of the c-extremization procedure \cite{Sen:2005wa,Kraus:2005vz}.

As an example of the procedure described above let us consider a theory of pure gravity with no matter fields $\phi_i$.  Diffeomorphism invariance implies that the action will be of the form
\be
\label{action}
S = {1\over 16 \pi G} \int \sqrt{g} \left(R - 2 \Lambda + c_1 R^2 + c_2 R_{\mu\nu} R^{\mu\nu} + c_3 R_{\mu\nu\rho\sigma}R^{\mu\nu\rho\sigma} +\dots + c_{cs} \Omega_{CS}\right)\;.
\ee
where we include the possibility of a gravitational Chern-Simons term in odd dimensions.
The stress tensor $T_{\mu\nu}$ is therefore function of the Riemann tensor and its covariant derivatives.  Because these are constructed out of the \sch{z,L}{d+3} metric, the stress tensor will automatically be invariant under the continuous \sch{z,L}{d+3} symmetries.  By general covariance, it will be conserved as well.  We conclude that in general \sch{z,L}{d+3} will be a solution to the equations of motion for arbitrary values of the coefficients $(c_1,c_2, c_3, \dots, c_{cs})$ appearing in the action in addition to possible zero parameter constraints (i.e. inequalities).  If we wish to fix a particular value of $z$ then we must impose an additional one parameter constraint.

We should make an additional comment on the higher curvature actions described above.  It is well known that terms in the action which are quadratic in the Riemann tensor may be shuffled around or removed by a field redefinition of the form
\be \label{fr}
g_{\mu\nu} \to \phi(R) g_{\mu\nu} + \psi(R)R_{\mu\nu}+\dots
\ee
However, this procedure mixes different orders in the derivative expansion, so
will not necessarily map solutions of the equations of motion to solutions of the equations of motion. In particular, a solution where different terms in the derivative expansion are balanced against one another will not necessarily be mapped to a new solution to the equations of motion.  To see that this is the case, let us imagine applying this change of variables (\ref{fr}) to the usual Einstein-Hilbert action.  The resulting equation of motion is fourth order rather than second order, so the number of solutions to the equations of motion is not preserved.   This subtlety is normally irrelevant, because the derivative expansion is viewed as a perturbative expansion in the sense of effective field theory.  In this case the perturbation series can not be trusted for solutions in which different orders in the expansion are balanced against one another.   Thus the classical solutions may be taken seriously only if one has an additional argument that higher terms in the perturbation series can be neglected.  

Notably, this can sometimes happen in supersymmetric systems, where higher order terms are expected to be absent\cite{Hanaki:2006pj}; an example where this is the case will appear in \cite{alex}.  The suppression of corrections beyond some fixed order can also happen via fine tuning, as for example in Banks-Zaks fixed points in which $N_{f}$ and  $N_{c}$  may be tuned so that precisely two orders compete but no higher-order contributions matter\footnote{We thank Shamit Kachru for discussions on this point.}.

\subsubsection{On the Non-Renormalization of AdS}
Before moving on, it is entertaining to rephrase our results in the special case $z=1$.  In this case, the conformal algebra may be extended to the full AdS algebra, $SO(d+1,2)$.  Remarkably, the only symmetric two-tensor invariant under this enlarged symmetry group is the AdS metric itself,
\be
\tau=\tau_{0} \[ {-dt'^{2} +d\vec{x}^{2}+dr^{2}\over r^{2}}\],
\ee
where $\tau_{0}$ is the only free parameter (note that we have changed coordinates away from light-cone to canonical Poincare coordinates).  The only possible effect of modification of the action on the Einstein equation is thus to shift the value of $\tau_{0}$, and thus to rescale the AdS radius, $R_{AdS}$.  Aside from this rescaling, the form of the AdS metric is not renormalized by quantum corrections to the effective action; this is the familiar statement that Anti-de Sitter space is a ``universal" solution of quantum gravity.

\subsection{A Toy Example: \sch{z}{d+3} Solutions of $R^{2}$ Gravity}
As an example of the above discussion, we consider the special case of pure gravity in $D=d+3$ dimensions with general quadratic corrections to the Einstein-Hilbert action,
\be\label{HCL}
{1\over 16 \pi G}  \int d^{D} x~~ \sqrt{-g}\left(R-2\Lambda+c_1 R^2 + c_2 R_{\mu\nu} R^{\mu\nu} + c_3 R_{\mu\nu\rho\sigma}R^{\mu\nu\rho\sigma}\right)\,,
\ee
and look for a Schr\"odinger solution of the form (\ref{SchMetric}).  According to the above analysis, this metric should be a solution for generic values of the parameters $c_i$ and $\Lambda$ in our Lagrangian, with $z$ and $L$ fixed in terms of these parameters.  The equations of motion for this theory take the form \cite{kp}
\be
R_{\mu\nu}-{1\over 2} R g_{\mu \nu}+\Lambda g_{\mu\nu}={1\over 2} T^{R}_{\mu\nu}\,,
\ee
where $T^{R}_{\mu\nu}$ is the effective stress tensor generated by the quadratic curvature terms (see the appendix for details).
Evaluated on our metric and expanded in the invariant tensors, $T^{R}_{\mu\nu}$ reduces to
two coefficients $\a^{T}$ and $\b^{T}$ which are unenlightening\footnote{For example, $\b^{T} = {(d-1)(d-4)\over L^{2}}(c_1 d(d-1)+c_2 (d-1)+2 c_3 )$.} but easily computable functions of $z$, $L$ and the $c_{i}$.
The equations of motion set these coefficients equal to those in (\ref{LHScoef}).  Keeping $z$ fixed, we can solve one of these equations by fixing $L$ in terms of the $c_i$ and $\Lambda$, with the second equation becoming a single condition on the parameters $c_{i}$ and $\Lambda$.

For example, consider the case $d=5$ with $\CL_{R^{2}}=(c~\! W^2-a~\!  GB)$, where $W^2$ and $GB$ are the Weyl-squared and Gauss-Bonnet terms, respectively. (This corresponds to $c_1=c/6-a$, $c_2=4a-4c/3$ and $c_3=c-a$.)  Solving the equations of motion fixes $z$ and $L$ in terms of $c$, $a$ and $\Lambda$ as,
\bea
z &=& \pm 1\,, \qquad L=\sqrt{-{3\pm \sqrt{9-12 a \Lambda}\over \Lambda}}\,,\\
z&=& \pm \sqrt{{8 c \Lambda+(9-12 a \Lambda)\pm 3 \sqrt{9-12 a\Lambda}\over 32 c \Lambda}}\,, \qquad L= \sqrt{- {3\pm \sqrt{9-12 a \Lambda}\over \Lambda}}\,.
\eea
Notice that AdS with $z=1$ is always a solution for any value of $a,c$. Also note that pure Gauss-Bonnet with $c=0$ leads to degenerate solutions. The interesting (perilous) solutions are those in the second line. Expanding around small $a,c$, we have two possibilities
\bea
z &=& \pm {3\over 4 \sqrt{c \Lambda}} + O(\sqrt{c},\sqrt{a})\,, \qquad L=\sqrt{-{6\over \Lambda}}+ O(a,c)\,,\\
z&=& \pm {1\over 4} \sqrt{4-{3a\over c}} + O(a,c) \,, \qquad L=\sqrt{-2 a}+O(a^{3/2},c^{3/2})\,.
\eea
The solution in the first line corresponds to a large $z$ since in our parametrization $\Lambda\sim O(1)$ while $c\sim O(1/N)$ as we explain in the appendix. If we demand that $z\sim O(1)$ then $c\sim O(1)$ and 
the approximation becomes suspect. 
The solution in the second line can in principle correspond to arbitrary $z$ by tuning $c,a$ but it generically corresponds to large curvature and hence the approximation breaks down.

Of course, these solutions only exist by balancing terms in the action which are linear and quadratic in the curvature, raising the obvious concern that truncating to quadratic order is not reliable.  Indeed, it is easily verified that even higher order terms will lead to order one corrections to the dynamical exponent.  We can avoid this danger by re-introducing a non-trivial matter sector: this introduces a third term to the equation of motion which allows us to make the contributions from the higher-curvature terms controllably small.  We will exploit this possibility in the next section by studying a particular matter sector, Type IIB string theory.  First, though, we show that our higher-curvature Lagrangian contains another non-relativistic geometry as a solution, the Lifshitz spacetime.

\subsection{Solutions with \lif{z}{d} Symmetry}

In this section we describe the construction of gravitational solutions with \lif{z}{d+2} symmetry.  We will follow the general strategy of the previous sections.
The \lif{z}{d+2} symmetry algebra is generated by linear momenta $P_i$, a Hamiltonian $H$ and angular momenta $M_{ij}$ enjoying the usual commutators, along with a dilatation operator $D$ with nonvanishing commutators
\be
[D,P_i] = i P_i,~~~~~[D,H] = i z H\;.
\ee
Following Kachru, Liu and Mulligan \cite{KLM}, we may realize these as isometries of a $d+2$ dimensional spacetime
\be \label{LifMetric}
ds^2 = L^2 \left( - {dt^2 \over r^{2 z}} + {d\vec{x}^2 + dr^2 \over r^2}\right).
\ee
which we refer to as \lif{z}{d+2}.
It is easy to check that the Killing vectors
\be\label{killings2}
M_{ij} = -i( x^i \p_j - x^j \p_i ),~~~~
P_i=-i \p_i,~~~~
H = -i \p_t,~~~~
D = -i (zt\p_t + x^i \p_i + r\p_r)\;,
\ee
generate the \lif{z}{d+2} algebra.  This observation led the authors of \cite{KLM}  to conjecture that quantum gravity in \lif{z}{d+2} is holographically dual to a quantum fields theory sitting at a quantum critical point with Lifshitz symmetry. Such critical points with $z = 2$ are well known in the condensed matter literature \cite{criticality}.

The general analysis of this spacetime proceeds exactly as above.  The most general invariant symmetric two-tensor $\tau = \tau_{\mu\nu} dx^\mu dx^\nu$ is
\be
\tau =  \alpha {dt^2 \over r^{2 z}} + \beta {d\vec{x}^2\over r^2} + \gamma {dr^2 \over r^2}+\delta {2 dr dt \over r^{z+1}}\;.
\ee
Conservation of the stress tensor $\nabla^\mu \tau_{\mu\nu}=0$ implies that
\be
\tau=  \alpha {dt^2 \over r^{2 z}} + \beta {d\vec{x}^2\over r^2}
+\left({(d-2) \beta - z \alpha \over d-2+z}\right) {dr^2 \over r^2}.
\ee
Thus there is a two parameter family of conserved stress tensors, just as in the Schr\"odinger case.

The metric (\ref{LifMetric}) contains two free parameters -- $z$ and $L$ -- just as in the Schr\"odinger case.  Following the logic of section 2.1, we therefore conclude that \lif{z,L}{d+2} may be obtained as a solution without any fine-tuning of the parameters in the Lagrangian.   
Likewise, we conclude that quantum corrections to the effective action will not spoil the existence of a Lifshitz invariant spacetime, although they may renormalize the parameters $L$ and $z$.

Like the Schr\"odinger metric, the Lifshitz metric may be found as the solution of a suitably-tuned higher-curvature Lagrangian.  For example, consider the Lagrangian studied above, (\ref{HCL}), with $c_{2}=c_{3}=0$, \ie\ pure $R^{2}$ gravity.  Evaluating the equations of motion on the Lifshitz metric leads to the relations\footnote{Note that while $AdS$ remain a solution of this higher-curvature gravity Lagrangian, our Lifshitz solution is on a separate branch of solutions. As with \sch{z}{d+3}, we will improve upon this spherical cow in the next section, where a non-trivial matter sector will give our constructions real legs.}
\be
c_{1} = {L^{2}\over 6+4z+2z^{2}}\;, ~~~~~~~~  \Lambda  = {3+2z+z^{2} \over L^{2}}\;.
\ee

This solution has the frustrating property that when the space is large, so must be the higher-curvature corrections, and vice versa.  So we appear forced outside the regime of validity of our perturbative description.
Indeed, these solutions balance curvature terms against curvature-squared terms, so we have no right to expect the quadratic approximation to be reliable.
To avoid this trap, we should re-introduce a non-trivial matter sector, for example the $p$-form action of \cite{KLM}.  We will study this system in the next section, where perturbatively small higher-curvature corrections will be shown to generate quantum corrections to the parameters $L$ and $z$ of the Lifshitz metric.

~

\section{Renormalizing $z$ in Type IIB and Other Animals}\scz

As we saw in the previous section, varying the coefficients of higher-curvature terms in the gravitational action shifts the dynamical exponent $z$ of \sch{z}{d+3} solutions of the unperturbed Lagrangian.  In this section we will explore this effect in \sch{z}{5} solutions of IIB supergravity, where higher curvature corrections to the bulk action will renormalize the dynamical exponent $z$ of the dual NRCFT.  This leads to explicit NRCFT-gravity dualities for general non-integral values of $z$.

We begin by considering the case of the five dimensional solution \sch{2}{5} originally presented in \cite{original} and embedded in IIB string theory in \cite{malda,Adams,Herzog}.  Incorporating higher curvature corrections will renormalize the dynamical exponent away from its classical integer value.  We then apply similar reasoning to the Lifshitz system studied in \cite{KLM}, where we again find $z$ shifted away from its classical value.

\subsection{Embedding Non-Galilean Schr\"odinger Spacetimes in Type IIB}
We start with the consistent truncation of type IIB supergravity on a Sasaki-Einstein 5-manifold described in \cite{malda}.
In this reduction, the only fields turned on are the metric, dilaton, five-form flux and the NS-NS three-form flux. Two different ansatz for the ten dimensional fields were presented in \cite{malda}. One of them gives rise to the $z = 2$ exponent, and the other to $z = 4$. We will concentrate on the case of $z = 2$,  although our arguments work just as well for $z = 4$. The Kaluza-Klein reduction leads to the five-dimensional action
\bea \label{Smalda} S &=& \frac{1}{2} \int d^5 x \sqrt{-g} \left[R + 24 e^{-u - 4 v} - 4 e^{-6u-4 v} - 8 e^{-10 v} - 5 (\nabla u)^2 - \frac{15}{2} (\nabla v)^2 \right. \nonumber \\
&&\left.- \frac{1}{2} (\nabla \Phi)^2 - \frac{1}{4} e^{-\Phi + 4 u + v} F_{\mu \nu} F^{\mu \nu} - 4 e^{-\Phi - 2 u - 3v} A_\mu A^\mu\right]\;,\eea
where $\Phi$ is the dilaton, and $u, v$ are related to the warp factors of the compactification. The gauge field $A$ is related to the B-field in ten dimensions. The action (\ref{Smalda}) admits solutions with all scalars set to zero so long as $F_{\mu \nu} F^{\mu \nu} = 0$ and $A_\mu A^\mu=0$ when evaluated in the solution. One can easily show that this is case for any Schr\"odinger spacetimes since the gauge field takes form $A \sim \a r^{-z} {dt}$.  The equations of motion that follow from (\ref{Smalda}) thus lead to the Schr\"odinger metric (\ref{SchMetric})
supported by vector dust,
\be 
A   = \a  {dt\over r^{z}} ,
\ee
with the coefficients fixed to
\be
L = 1\;, ~~~~~~~~~~ z=2\;, ~~~~~~~~~~ \a=1\;.
\ee

Let's now add general $R^{2}$ corrections to the five dimensional effective Lagrangian (for a concise discussion of the origin of such terms in the IIB effective action, see the appendix),
\bea \label{S5}
\CL = R + 12 - \frac{1}{4} F_{\mu \nu} F^{\mu \nu} - 4 A_\mu A^\mu
+ c_1 R^2 + c_2 R_{\m\n} R^{\m\n} + c_3 R_{\m\n\r\s}R^{\m\n\r\s} + \ldots
\eea
Note that we have again set all scalars to zero -- this is consistent in the zero temperature case on which we focus in this section, though must be relaxed when considering black hole solutions. Note too that we have suppressed other $\a'^{2}$ corrections which depend on the gauge flux (\eg\ $F^4$, $(dF)^{2}$, $RF^2$).  However, since\footnote{This makes the analysis quite similar to the argument in \cite{mps} where it was shown why the five-form flux could not affect the finite coupling result for $\eta/s$ in the AdS-Schwarzschild case.} $F^{2}$=0 and $dF$=0, the only terms we need to worry about are of the form\footnote{A topological term like $A\wedge R \wedge R$ could also in principle be present at this order but one can explicitly check that $R\wedge R=0$ so that the gauge field equations of motion are not affected.} $RF^{2}$. An explicit calculation with all possible tensor contractions shows that this will not affect the result for the dynamical exponent (we will however return to this term when studying the matter EOMs).  With these simplifications, the equations of motion that follow from this action are:
\bea
\label{eq1}
\nabla_{\mu} F^{\mu \nu} &=& 8 A^{\nu} \;,\\
\label{eq2}
G_{\mu\nu}-6 g_{\mu\nu}&=& \frac{1}{2} F_{\mu \sigma} F_\nu^{\;\;\sigma}  + 4 A_\mu A_\nu  - \frac{1}{2} g_{\mu \nu} \left(\frac{1}{4}F^2+ 4 A^2\right)+{1\over 2} T^{R}_{\mu\nu}\,,
\eea
where $T^{R}_{\mu\nu}$ is as described in Section 2 and presented in the appendix.

Since the $c_i$ represent $\alpha'^2$ corrections, it is reasonable to treat these corrections perturbatively and look for deformed solutions order by order in the $c_i$, at least so long as $R_{AdS} \gg \ell_{s}$. By the arguments in Section 2, the worst that can happen to the metric under such a perturbation of the action is a renormalization of the parameters $L$ and $z$.
Expanding Eqs. (\ref{eq1})  and (\ref{eq2}) to linear order in $c_i$'s and the above perturbations, we indeed find a renormalized solution. Defining $c_4=5c_1+c_2$ as in the appendix,
\bea
\delta z &=& -\frac{8}{27} (2 c_4+c_3)\;,\\
\delta L &=& -\frac{3}{27} (2c_4+c_3)\;,\\
\delta\a &=& \frac{5}{27} (-122 c_1+8 c_2+139 c_3)\;.
\eea
$R^{2}$ corrections to the IIB action thus renormalize the dynamical exponent away from its classical value, with the precise value depending on what corrections are generated.

At this point, you might be worrying that the value of $z$ appears dangerously coordinate dependent, since the $R^{2}$ terms in the gravity action (and thus their coefficients, the $c_{i}$) may be shuffled around by a field redefinition $g_{\m\n}\to g_{\m\n}+f(R)R_{\m\n} ~ ...$.  But $z$ is a physical, observable quantity in the NRCFT -- it controls the dispersion relation, $\w\sim k^{z}$ -- so it had better remain invariant under field redefinitions!  It is thus reassuring to note that the specific combination appearing before us, $2c_{4}+c_{3}$, is in fact invariant under field redefinitions of the metric.  Sound the trumpets.

Actually, not just yet on the trumpets.  Of the parameters defining our solution, two depend on the invariant combination $2c_{4}+c_{3}$.  However, $\a$, the density of the dust, appears to depend on a different, field-redefinition-dependent, combination of the $c_{i}$.  In fact, under this field-redefinition of the metric, the $F^{2}$ terms themselves get corrected, with e.g. new $F^{2}R$ terms generated in the action (notably, we suppressed terms of this form above).  It is straightforward to check that the effect of these terms is only to shift $\a$, not $z$ or $L$.  A careful treatment of all such terms necessarily gives a coordinate-independent result.

\subsubsection{Gauge Theory Duals with $z\neq2$}

At this point, we have constructed solutions of higher-curvature modified IIB string theory which realize the Schr\"odinger group as their isometry group.  We have yet to answer, however, a basic question: who puts the ``gauge'' in our gauge-gravity duality?

As a warm-up, recall the answer in the special-conformal case, $z=2$.  In that case, the duality could be understood as mapping IIB on the Schr\"odinger spacetime to a twisted \bDLCQ\ of $\CN=4$ with gauge group $SU(N)$.  Concretely, the \bDLCQ\ is defined by choosing a light-like direction, $x^{+}$, and an $R$-symmetry generator, $J_{R}$, and demanding that all the fields of the gauge theory be periodic along $x^{+}$ up to an $R$-charge dependent phase,
\be
\Phi(x^{+}+L) = e^{-i\b q_{R}} \Phi(x^{+}),
\ee
where $\Phi$ stands in for any of the fields of the theory and $q_{R}$ its charge under the chosen $R$-current, $J_{R}$.  Expanding in modes along the compactified $x^{+}$ direction, the modified periodicity condition becomes,
\be
\Phi=\sum_{\ell} e^{i \ell x^{+} }\Phi_{\ell} ~, ~~  \ell  = {2\pi\over L}n - \b q_{R}~,~~n\in\IZ ~.
\ee
The effect of non-zero $\b$ is thus to shift the moding by a constant proportional to the $R$-charge. To see that the resulting theory is effectively non-relativistic, note that the kinetic terms for, say, a scalar $\phi$ become,
\bea
(\p\phi)^{2} &=& -2~\! \p_{+}\phib \p_{-}\phi + (\vec{\nabla}\phi)^{2} \non\\
&=& \sum_{\ell} \[-2i\ell ~\! \phib_{\ell} \p_{-}\phi_{\ell} + (\vec{\nabla}\phi_{\ell})^{2}\] .
\eea
Iff $\b\neq0$, every mode of $\phi$ has a non-trivial first-order kinetic term\footnote{If $\b=0$, $\Phi$ picks up a zero mode without a kinetic term, indicating a strong-coupling break-down of the non-relativistic Kaluza-Klein reduction along the light-like $x^{+}$ circle and reminding us that the theory is fundamentally relativistic.  Of course, even with $\b\neq0$, any field with trivial $R$-charge will also have a badly-behaved zero mode. In the cases of interest to us, the only such field is the gauge boson whose zero mode enforces a Gauss law.  While this is an important feature of our theories, for our concerns in this paper, where most computations are conducted on the gravity side, it can safely be swept under the rug.}.  The action is thus invariant under the non-relativistic Galilei group, with $x^{-}$ playing the role of time.  At zero coupling, the entire action is in fact invariant under the full $z=2$ Schr\"odinger group with $N=\p_{x^{+}}$.  That $\CL$ remains conformal at finite coupling is a remarkable and beautiful fact.

Let's now return to our $R^{2}$-modified examples with $z\neq2$.  To whom are {\em they} dual?  Clearly the answer depends on precisely what combination of higher-curvature terms are present in the IIB effective action.  As reviewed in the appendix, one well-understood and controlled example involves adding D7 branes to a stack of $N$ D3 branes at an orientifold fixed plane and taking the decoupling limit.  The worldvolume theory is then an $\CN=2$ $Sp(N)$ gauge theory with $n_{H}=2N^{2}+7N-1$ hypermultiplets and $n_{V}=2N^{2}+N$ vector multiplets.  A standard argument relates the trace anomaly of this theory to the coefficients of the Weyl and Gauss-Bonnet terms in the dual gravitational action.  But $\CN=2$ ensures that the trace anomaly takes a very simple form in terms of $n_{H}$ and $n_{V}$.  This allows us to fix the coefficient of the $R^{2}$ terms in the gravitational action in terms of the rank, $N$, of the dual $Sp(N)$ gauge theory.

Now, as we have seen, such an action still has AdS solutions with $z=1$ and slightly modified $L$.  In fact, this is completely robust: the AdS isometry group only preserves a single rank-2 symmetric tensor, so any deformation of the gravitational action may be absorbed in a rescaling of the AdS radius, $L$.  Other features of these solutions, of course, {\em do} depend on the details of the $R^{2}$ corrections, leading to interesting physics; such solutions have thus received considerable attention.

As we have also seen, the curvature-corrected IIB action also includes \sch{}{5} solutions whose dual NRCFT is now easily identified as the \bDLCQ\ of $\CN=2$ $Sp(N)$ SYM with $n_{H}(N)$ hypers and $n_{V}(N)$ vectors.  In terms of the gauge theory parameters (\ie\ expressing the $c_{i}$ in terms of $N$; for details, see the appendix), our strong-coupling gravitational result for the renormalized dynamical exponent becomes,
\be
z = 2+ \frac{2}{27} \frac{1}{N} + \CO\({1\over N^{2}}\),
\ee
with $z\to2$ in the classical large-$N$ limit.  Matching this prediction on the field theory side would be a remarkable check, so let's see what we can say.

Classically, \ie\ at zero coupling, the kinetic terms for all fields but the vector zero modes take the Galilei-invariant form described above, so we expect $z=2$.  This matches the $\CO(N^{0})$ result on the gravity side nicely.  Once we turn on interactions, of course, we can no longer trust the classical result; in general, $z$ will be renormalized away from its classical value.  Precisely this effect was studied by Son and by Sachdev and Nikolic \cite{N}, among others, who found that, eg in a system of fermions respecting a global $Sp(N)$ symmetry at unitarity, critical exponents generally receive non-trivial $1/N$ corrections\footnote{For example, in an $N$-species model of graphene, Son computed the dynamical exponent to be $z=1-{4\over\pi^{2} N}$, while in a global $Sp(2N)$ model of unitary fermions, Sachdev and Nicolic computed the critical fermi energy to be ${\eps_{F}\over T}|_{T_{c}} \sim 2 +{5.3\over N}$.}.  In {\em our} finite-$N$ field theory, then, it is natural to expect a similar non-zero finite-$N$ renormalization of the exponent,
\be
z=2+\zeta {1 \over N} + \CO\({1\over N^{2}}\),
\ee
where the explicit coefficient, $\zeta$, is model-dependent but generically non-zero.  

This qualitative agreement suggests a very strong and concrete test of our proposal.  It should be possible to extend these large-$N$ techniques to our gauge theories, and to use them to compute $z$ in $1/N$ expansion.  Comparison to our gravitational prediction would provide a very sharp test of the strong, \ie\ quantum, version of this non-relativistic gauge-gravity correspondence, which predicts
\be
\zeta={2\over 27}.
\ee
It should be noted, however, that this is a rather non-trivial computation, which we pose as a very much open problem.

\subsubsection{Higher Curvature Corrections?}
So much for the $R^{2}$ terms in the IIB action.  What about $\alpha'^3$ corrections? Our 5d IIB action receives just such terms from the well-known $C^4$ term in 10d. The non-zero components of the Weyl tensor are
\be
C_{tr tr}={2L^2 (2z-1)(z-1)\over 3 r^{2(z+1)}}\,,\quad C_{tx_1 tx_1}=C_{tx_2 tx_2}=-{L^2(2z-1)(z-1)\over 3 r^{2(z+1)}}\,,
\ee
where we are now considering a general dynamical exponent.
The $\alpha'^3$ correction in IIB is proportional to $\int \sqrt{-g} W$ where $W$ is given by
\be
W=-\frac 12  C_{a b c d} C^{a b}_{\ \ \ e f} C^{c e}_{\ \ \ g h} C^{d g f h}+C_{a b c d} C^{a \ c}_{\ e \ f}C^{b \ e}_{\ g \ h} C^{d g f h} \,.\label{C4}
\ee
If one assumes a solution of the type $A_5\times M_5$ where $M_5$ is a 5d-compact manifold, then it can be shown that after a field redefinition \cite{amps}, the correction in the effective theory in 5-dimensions is again given by (\ref{C4}) with the indices now restricted to lie in the non-compact directions. It can be easily verified that $\delta W/\delta g^{\mu\nu}$ vanishes. On varying one of the Weyl tensors, the remaining 4 index $C^3$ has to be evaluated on-shell and this vanishes. Hence there are no $C^4$ modifications to this metric. It is possible that the flux terms at this order may play a role; we leave the analysis of this possibility as an open problem.

What about still higher-orders?  Recalling our classification of symmetric 2-tensors in Section 2, it should be clear that, up to the assumption that the matter sector is sufficiently general to allow solutions of the full equations of motion, additional higher-curvature corrections can do nothing but further re-normalize our solutions.  So long as we begin with a solution with curvatures well below string or Planck scale, these renormalizations should be smaller still.

Finally, it is useful to observe that the results of this section can readily be generalized to any dimensions and for a general matter sector, and thus well beyond the 5d IIB effective action.
Rather then belabor this point, however, let's look at a related non-relativistic system, the Lifshitz spacetime.

\subsection{Non-Galilean Lifshitz Spacetimes in 4d Gravity}
Another interesting class of non-relativistic metrics with time-reversal invariance is the Lifshitz metric, (\ref{LifMetric}), first presented by Kachru, Mulligan and Liu in \cite{KLM}.  This metric solves the EOM that follow from the effective action, 
\def\c{\g}
\bea \label{KMLAction}
S &=& \int d^4 x \sqrt{-g} \left(R - 2 \Lambda \right) -\frac{1}{2} \int \left( F_2 \wedge * F_2 +   F_3 \wedge * F_3\right) - \c \int B_2 \wedge F_2\;, \eea
where $F_3 = d B_2$ and $F_2 = d A_1$. The EOM are
\be  \label{Feq} d * F_2 = -\c F_3\;,\;\;\; d*F_3 = \c F_2\;,\ee
and
\be G_{\mu \nu} + \Lambda g_{\mu \nu}  = \sum_{p = 2,3} \frac{1}{2 p!} \left( p F_{\mu \rho_2 \cdots \rho_p} F_\nu^{\; \;\rho_2 \cdots \rho_p}
- \frac{1}{2} g_{\mu \nu} F_p^2\right)\;.\ee

The solution to these EOMs takes the form:
\be F_2 = A \theta_r \wedge \theta_t\;,\;\;\; F_3 = B \theta_r \wedge \theta_x \wedge \theta_y\;,\ee
where
\be \theta_t = L  \frac{dt}{r^z} \;,\;\;\; \theta_{x^i} =  L  \frac{dx^i}{r} \;,\;\;\; \theta_r = L \frac{dr}{r}\;,\ee
~
\be\label{eqs1} L^2 = - \frac{z^2 + z + 4}{2 \Lambda}\;,\;\;\; A^2 = \frac{2 z(z-1)}{L^2}\;,\;\;\; B^2 = \frac{4 (z - 1)}{L^2}\;,\ee
and
\be  \label{zeq} 2 z = (\c L)^2\;.\ee

Note that the value of the coupling $\c$ and the cosmological constant $\Lambda$ determine the dynamical exponent. To make this explicit, one can use the first equation in (\ref{eqs1}) in the consistency condition (\ref{zeq}). We then get a quadratic equation which can be solved for $z$ as a function of $\c$ and $\Lambda$.  To simplify our lives, let's fix units so that $\Lambda$ takes the  conventional value of the cosmological constant,
$\Lambda = -\frac{1}{2} (D- 2)(D- 1) = -3$,
where the last equality assumed we are working in $D=4$. The dynamical exponent is then,
\be z_\pm = \frac{12-\c^2\pm\sqrt{3} \sqrt{-5 \c^4-8 \c^2+48}}{2 \c^2}\;.\ee The two roots correspond to dynamical exponents greater ($z_+$) and less ($z_-$) than $z = 2$. The ``critical" value of $z = 2$ is obtained when $\c^2 = 12/5$. The value of $L, A, B$ are then determined in terms of $z$ or $\c$.

How does the dynamical exponent get modified by the inclusion of higher derivative corrections to the action (\ref{KMLAction})? Explicitly, consider the action
\be \label{KMLplus} S_{KLM}' = S_{KLM} + \int d^4 x \sqrt{-g}\(c_1 R^2 + c_2 R_{\m\n} R^{\m\n} + c_3 R_{\m\n\r\s}R^{\m\n\r\s}\)\;,\ee
where the $c_{i}$ are again assumed to be small.
Let $z_0$ be the ``classical" value of the dynamical exponent obtained without the higher derivative corrections (determined by $\c$ from $\c = 12 z_0/(z_0^2+z_0+4)$) and write $z = z_0 + \delta z$; similarly, let $A = A_0 + \delta A$, with $A_0^2 = 12 (z_0-1) z_0/(z_0^2+z_0+4)$.
Note that the EOM for the gauge fields is unchanged by the addition of the higher curvature terms, though the coefficients $A$ and $B$ and the one-forms $\t_{\mu}$ will themselves be modified,
\be \theta_t = L \frac{dt}{r^{z_0  + \delta z}} \;,\;\;\; \theta_{x^i} =  L \frac{dx^i}{r}\;,\;\;\; \theta_r = L \frac{dr}{r}\;,\ee
and
\be L^2 = \frac{2}{\c^2} (z_0  + \delta z), ~~~~ B = - A \frac{2}{\c L} \;.\ee
Solving the linearized EOM then gives,
\be 
\delta z ={8z_{0}(z_{0}-1) \[3c_{4}(z_{0}^{2}+2z_{0}+3)+c_{3}(9z_{0}^{2}-22z_{0}+7)\] \over 5 (z^{2}_0-4)(z_{0}^{2}+z_{0}+4)}
\ee
and
\be
\delta A_0 =
{\sqrt{3}(z_0^2+z_0-4) 
\over
\sqrt{z_{0}(z_{0}-1)(z_{0}^{2}+z_{0}+4)}
} \d z
\ee
Notably, these results are badly behaved near $z_{0}=1,2$, so these cases of must be treated separately. As expected, the $AdS$ case $z_0 = 1$ does not receive any corrections. Somewhat surprisingly, the case $z_0 = 2$ seems to be protected too, at least at this order in perturbation theory.  However, we see no sign of an extra symmetry protecting this solution -- indeed, it is easy to construct ad-hoc higher-curvature terms which do appear to renormalize $z_{0}=2$.  It thus seems unlikely to us that this solution remains protected at higher orders in the full quantum theory.

Unfortunately, we do not as yet know how to embed these solutions in string theory (which makes the problem of possible higher-order corrections extremely open-ended), or to identify unambiguously the dual NRCFT.  Nonetheless, whatever the dual theory, we expect that renormalization of its dynamical exponent is dual to the renormalization of the gravitational action by higher-curvature terms.

\section{Violating The KSS Bound in an NRCFT}\scz
%

We can now apply our machinery to a test of the KSS conjecture in non-relativistic CFTs.  In principle, the way to proceed is to heat up the zero-temperature NRCFT by holding its feet to the Hawking fire.  More precisely, the partition function for the NRCFT in a thermal ensemble may be found by evaluating the gravitational action on a suitable black-hole solution and exponentiating.  Computing the viscosity-to-entropy ratio is then a straightforward but intricate application of real-time gauge-gravity duality prescription.

Unfortunately, finding the black hole solution of our higher-curvature theory is far more difficult than the zero temperature case above, since the form of the solution is no longer fixed by Schr\"odinger invariance.  The in-principle-straightforward computation of $\eta\over s$ is thus, in practice, analytically quite challenging.

Happily, we can use a few simple facts about our Schr\"odinger spacetimes to derive the result without any messy mucking about with numerics.  Look back at the Schr\"odinger metric, (\ref{SchMetric}).  For any $z>1$, the $dt^{2} / r^{2z}$ term falls off more rapidly than the remaining light-cone-AdS piece of the metric as we approach the zero-temperature horizon at $r\to\infty$.  The geometry near the horizon thus reduces to (a DLCQ of) pure AdS with $R_{\rm AdS}=L$.  Puffing up the zero-temperature horizon, $r\to\infty$, to a fluffy finite-$T$ horizon should thus lead to a black hole whose near-horizon geometry is the same as the usual AdS black hole.  Indeed, this is precisely what was found by explicit construction in the $z=2$ and $z=4$ cases studied in \cite{Adams,Herzog,malda}.  This is also the structure found in other families of asymptotically-Schr\"odinger black holes \cite{AdamsWIP}.  As beautifully argued in \cite{Iqbal:2008by}, we may thus import the computation of the viscosity-to-entropy ratio directly from the case of the asymptotically-AdS black hole dual to the $Sp(N)$ theory at finite temperature \cite{kp}\ to get\footnote{It may be tempting to write the RHS of this relation in terms of $z-2$.  This would be misleading -- it is not the dynamical exponent departing from $z=2$ that modifies the KSS bound, as is clear from the fact that the same relation holds for the $z=1$ AdS case.  Rather, both $z$ and $\eta\over s$ (or rather, as emphasized in \cite{Iqbal:2008by}, $G_{N}$) receive quantum renormalization, with the relation between $\eta\over s$ and $z$ fixed by non-universal features of the system.},
\be
{\eta \over s} = {1\over 4\pi}\(1-{1\over 2N}\) \;.
\ee
Indeed, following Iqbal and Liu, we may import all of the universal hydrodynamics of our system directly from the AdS result.  It would thus be very interesting to go beyond this universal hydrodynamic sector.  For that, however, we need a full black hole solution; we leave this thorny problem to future work.

\section{Conclusions}\scz

We have argued that higher-curvature corrections to the gravitational action renormalize the dynamical exponent $z$  of a Schr\"odinger solution to Einstein's equations.  This observation allowed us to embed a class of fractional-$z$ Schr\"odinger spacetimes in string theory, identify the non-relativistic conformal field theories which we conjectured to be dual to these spacetimes, and begin studying their properties at strong coupling.  For example, we deduced the ratio of viscosity to entropy density $\eta / s$ for a simple class of NRCFTs with non-integral $z$.  This ratio varied with $z$ in a perturbatively computable way.  Notably, this result was derived indirectly; an explicit computation awaits the construction of asymptotically Schr\"odinger black hole solutions of the higher curvature gravity theory.  That said, our indirect argument relies on the Hong-Iqbal proof of universality, whose assumptions .  So this indirect argument is an aesthetic lacuna, not a technical one.

In our discussion of finite temperature effects, there are several important points to note.  First, while such a symmetric AdS black hole embedded in asymptotically Schr\"odinger space is certainly {\em a} finite-temperature solution, it is by no means clear that it is the {\em only} such solution.  Indeed, phase transitions in the boundary theory should correspond to instabilities of the bulk solution\footnote{For a discussion of this connection in AdS/CMT, see for example \cite{Phase} and references therein. We thank J. McGreevy for illuminating discussion on this point.}, and it seems likely that the phase structure of these theories may sometimes be non-trivial.  Second, while we were not able to find an analytic solution for the the black hole metric with $z\neq2$, this problem should be numerically tractable.  Note that a numerical solution would likely also be sensitive to the stability of the solution, and thus shed light on the phase structure of the dual theory.  Work in this direction is in progress. 

Finally, it is interesting to ask what disaster, if any, might be signaled by a violation of the KSS bound.  It was shown in \cite{Steve}\ that $R^{2}$ corrections which strongly violate KSS bound $\frac{\eta}{s} < \frac{16}{25} \frac{1}{4\pi}$ led to the introduction of closed timelike curves\footnote{An equivalent bound deriving from unitarity of the dual field theory appears in \cite{Hofman:2008ar} -- it would be very interesting to identify any similar bound in the non-relativistic context.}.  Thus sufficiently strong violations of the KSS bound are not consistent with relativistic causality in the bulk.  Of course, since our system is already non-relativistic this can hardly be a concern.  However, the observations of \cite{AllanNima} would suggest that the problem with strong violations of KSS are as much about unitarity and locality as about causality, and should persist in the non-relativistic limit.  It would be interesting to further elucidate this mystery.

\vspace{1cm}
{\center \bf Acknowledgements\\}\noindent
We thank
Andrew Frey,
Shamit Kachru,
Joshua Lapan,
Albion Lawrence,
Hong Liu,
John McGreevy, 
Rob Myers,
Yusuke Nishida, 
Megha Padi,
Miguel Paulos,
Subir Sachdev,
Omid Saremi,
Ashoke Sen and
Andrew Strominger
for discussions.  
We also thank the 2008 BIRS Workshop on Emerging Directions in String Theory where this collaboration was initiated.
A.A. thanks the organizers and participants of the 2008 Aspen Center for Physics Workshop on String Theory where some of this work was presented, and the Sippewissett Institute for Advanced Study for hospitality during the writing of this paper.
The work of A.A. is supported in part by funds provided by the U.S. Department of Energy (D.O.E.) under cooperative research agreement DE-FG0205ER41360.
The work of A.M. is supported by the National Science and Engineering Research Council of Canada.
The work of A.S. and S.V. is supported by the Government of Canada through Industry Canada and by the Province of Ontario through the Ministry of Research \& Innovation.

\begin{appendix}

\section{$R^{2}$ Terms in the Action and $T^{R}_{\m\n}$}\scz

Consider the action,
\be
S = \frac{1}{\kappa^2}\int d^D x \sqrt{-g}(R +\Lambda + c_1 R^2 + c_2 R_{\mu\nu}R^{\mu\nu} + c_3 R_{\m\n\r\l}R^{\m\n\r\l})\,.
\ee
In terms of the $D$-dimensional Weyl and Gauss-Bonnet quadratic curvature invariants,
\bea
W&=&C_{\mu\nu\rho\lambda}C^{\mu\nu\rho\lambda}=R_{\mu\nu\rho\lambda}R^{\mu\nu\rho\lambda}-{4\over D-2} R_{\mu\nu}R^{\mu\nu}+{2\over (D-1)(D-2)}R^2\,,\\
GB&=& R_{\mu\nu\rho\lambda}R^{\mu\nu\rho\lambda}-{4} R_{\mu\nu}R^{\mu\nu}+R^2\,,
\eea
this becomes,
\be
S = \frac{1}{\kappa^2} \int d^D x \sqrt{-g}(R+\Lambda+\a_{1} GB +\a_2 W+\a_3 R^2)\,,
\ee
where
$c_{1}=\a_{1}+{1\over6}\a_{2} +\a_{3}$,
$c_{2}= -4(\a_{1}+{1\over3}\a_{2})$ and
$c_{3}=\a_{1}+\a_{2}$.

Varying with respect to the metric gives the ``stress tensor'',
\bea
T^{R}_{\mu\nu}&=&
c_3 (g_{\mu\nu}R^2-4 R R_{\mu\nu} +4 \nabla_\nu\nabla_\mu R-4 g_{\mu\nu} \Box R)\nonumber \\
 &+&
 c_2 (g_{\mu\nu} R_{\rho\sigma}R^{\rho\sigma} +4 \nabla_\alpha\nabla_{(\nu} R^\alpha_{\ \mu)}-2\Box R_{\mu\nu}-g_{\mu\nu}\Box R-4 R^\alpha_{\ \mu} R_{\alpha \nu}) \nonumber \\
 &+&
 c_1 (g_{\mu\nu}R_{\alpha\beta\gamma\delta}R^{\alpha\beta\gamma\delta}-4 R_{\mu\alpha\beta\gamma}R_{\nu}^{\ \alpha\beta\gamma}-8\Box R_{\mu\nu}+4 \nabla_{\nu}\nabla_{\mu}R+8 R^\alpha_{\ \mu}R_{\alpha\nu}-8 R^{\alpha\beta}R_{\mu\alpha\nu\beta})\,.\nonumber \\
 \eea

\section{$R^{2}$ Terms in IIB and their Gauge Theory Origins}\scz
The purpose of this appendix is to fix the $R^2$ terms in the action in terms of gauge theory variables. The simplest example \cite{blau} is the gravity dual to ${\cal N}=2$ superconformal field theory with $Sp(N)$ gauge group with four fundamental and one antisymmetric traceless hypermultiplet. The string theory dual is IIB on $AdS_5\times S^5/Z_2$ in which 8 D7 branes are coincident on an O7 plane. The D7 branes and O7 plane wrap an $S^3$ which is the fixed point locus of the $Z_2$. Using the Weyl anomaly and holographic recipes, the coefficients of the higher derivative $R^2$ terms in the 5-dimensional effective action can be determined \cite{blau, nd}. These are the $R^2$ terms that we will use in the calculations in section 3.
Let's focus on the case of 5d compactifications of Type IIB Supergravity, whose pure curvature terms take the form
\be
S = {1\over \kappa^2}\int d^5 x \sqrt{-g}\left(R +{12\over {\ell^2}} + \ell^2(c_1 R^2 + c_2 R_{\mu\nu}R^{\mu\nu} + c_3 R_{\m\n\r\l}R^{\m\n\r\l})\right)\,.
\ee
The trace anomaly for the boundary theory was derived holographically in \cite{blau,nd}\ to be,
\be
\kappa^2\langle T_m^m\rangle=\underbrace{(-{L^3\over 8}+5 c_1 \ell^2 L+ c_2 \ell^2 L)}_{\alpha_1}(GB_4-W_4)+\underbrace{{c_3 \ell^2 L\over 2}}_{\alpha_2} (GB_4+W_4)\,,
\ee
where $L$ is the AdS-radius.
Let's define $c_2+5 c_1 \equiv c_4$, as only this combination appears in what follows. We have \cite{nd}
\be \label{la}
2 L^2=\left({1 + \sqrt{1-{8\over 3} (2 c_4+ c_3)}}\right){\ell}^2\,.
\ee
The trace anomaly can also be written as
\be
\kappa^2\langle T_m^m \rangle = \left(2\alpha_2 R_{m n r s}R^{m n r s}-2(\alpha_1+3\alpha_2)R_{m n}R^{m n}+{2\over 3}(\alpha_1+2\alpha_2)R^2\right)\,,
\ee
where $m,n,r,s$ are 4-d indices. We know that for a ${\cal N}=2$ theory with $n_V$ vector multiplets and $n_H$ hypermultiplets, the trace anomaly is
\be \label{anom1}
\langle T_m^m \rangle =\tau \left((n_H-n_V) R_{m n r s}R^{m n r s}+12 n_V R_{m n}R^{m n}-{1\over 3}(11 n_V+n_H)R^2\right)\,.
\ee
Here \cite{nd} $\tau={1\over 24 \times 16 \pi^2}$. Comparing coefficients we get 3 equations for 2 variables,
\be
\tilde n_H-\tilde n_V=2\alpha_2\,,\quad 12 \tilde n_V=-2(\alpha_1+3\alpha_2)\,,\quad 11 \tilde n_V+\tilde n_H=-2 (\alpha_1+2\alpha_2)\,.
\ee
The tilde-d variables are the original ones times $\tau \kappa^2$.

Consider specifically the case when the gauge group is $Sp(N)$. In this case with $1/\kappa^2=N^2/(4 \pi^2 L^3)$ and,
\be
n_H=2 N^2+7 N-1\,,\qquad n_V=2 N^2+N\,,
\ee
using which we get perturbatively in $1/N$
\be
c_3={1\over 16 N}+O(1/N^2)\,,\qquad c_4=-{5 \over 32 N}+O(1/N^2)\,.
\ee
This is consistent with the findings in \cite{kp}. In the main text we sometimes set $L=1$ which to $O(1/N)$ is the same as setting ${\ell}=1$.

\end{appendix}

\end{document}